\documentclass[prb,superscriptaddress,twocolumn,amsmath,amssymb,floatfix]{revtex4-1}
\usepackage{latexsym}
\usepackage{dcolumn}
\usepackage[dvips]{graphicx}
\usepackage{amssymb}
\usepackage{graphicx}
\usepackage{psfrag, color}
\usepackage{verbatim}
\usepackage{amsmath}
\usepackage{epsf}
\usepackage{bigstrut}

\newcommand{\vk}{{\bf k}}

\newcommand{\ve}{{\varepsilon}}
\newcommand{\vq}{{\bf q}}
\newcommand{\tat}{{\tau_t}}
\newcommand{\tas}{{\tau_s}}

\begin{document}


\title{Carrier screening, transport, and relaxation in 3D Dirac semimetals}

\author{S. Das Sarma}
\affiliation{Condensed Matter Theory Center, Department of Physics, 
	 University of Maryland,
	 College Park, Maryland  20742-4111}
\author{E. H. Hwang}
\affiliation{Condensed Matter Theory Center, Department of Physics, 
	 University of Maryland,
	 College Park, Maryland  20742-4111}
\affiliation{SKKU Advanced Institute of Nanotechnology and Department
  of Physics, Sungkyunkwan
  University, Suwon 440-746, Korea } 
\author{Hongki Min}
\affiliation{Condensed Matter Theory Center, 
Department of Physics, University of Maryland, College Park,
Maryland  20742-4111}
\affiliation{Department of Physics and Astronomy, Seoul National University,
Seoul 151-747, Korea}


\begin{abstract}
A theory is developed for the density and temperature dependent
carrier conductivity in doped three-dimensional (3D) Dirac materials
focusing on resistive scattering from screened Coulomb disorder due to
random charged impurities (e.g., dopant ions and unintentional
background impurities). The theory applies both in the undoped
intrinsic (``high-temperature'', $T \gg T_F$) and the doped extrinsic
(``low-temperature'', $T \ll T_F$) limit with analytical scaling
properties for the carrier conductivity obtained in both regimes, where $T_F$ is the Fermi temperature corresponding to the doped free carrier density (electrons or holes). The
scaling properties describing how the conductivity depends on the density and temperature
can be used to establish the Dirac nature of 3D
systems through transport measurements. 
We also consider the temperature dependent conductivity limited by the acoustic phonon scattering in 3D Dirac materials.
In addition, we theoretically calculate and compare the single particle relaxation time $\tas$, defining the quantum level broadening, and the transport scattering time $\tat$, defining the conductivity, in the presence of screened charged impurity scattering.
A critical quantitative analysis of the $\tat/\tas$ results for 3D Dirac materials in the presence of long-range screened Coulomb disorder is provided.

\end{abstract}

\maketitle

\section{introduction}

Following the very extensive research activity in graphene over the
last 10 years \cite{geim2007,neto,peres,dassarma2011}, a great deal of
interest in condensed matter physics 
(and beyond \cite{tarruell2012}) has focused on other 2D and 3D Dirac
materials where the 
elementary low-energy noninteracting electronic energy dispersion is
linear and can be written in the continuum long wavelength limit as:
$E({\bf k}) = \pm \hbar v_F |{\bf k}|$ where $v_F$ is the (Dirac)
Fermi velocity and the $+/-$ signs denote electron/hole
(conduction/valence) bands, respectively, meeting at the so-called Dirac
point ${\bf k}=0$ in our notation (where the conduction/valence bands
meet making the system a gapless semiconductor or a semimetal). 
Considerable theoretical work focusing on bulk chiral materials with a gapless band structure including Weyl and 3D Dirac semimetals \cite{wan2011, singh2012,ohtsuki2014} has been carried out recently.
This linear gapless massless chiral Dirac band dispersion defines a solid
state Dirac material, and exciting 
new materials and experimental developments
\cite{liu2014,borisenko2014,neupane} have led to the laboratory 
realization of 3D Dirac materials (i.e., essentially the 3D
generalization of 2D graphene) by several groups. 
We mention here that for our purpose in the current work, Dirac and Weyl systems are equivalent (i.e. our work applies to both systems equally) as long as the appropriate degeneracy factor (called `$g$' in our work) is used in the calculations.  We will use the generic expression `3D Dirac system' to describe both materials in the current work.

An intrinsic or undoped 3D Dirac system is a semimetal with the linear electron (conduction)-hole (valence) bands touching at the Dirac point with the valence band completely full and the conduction band completely empty (at $T=0$) since the chemical potential or the Fermi level is precisely at the Dirac point.  Such a system is charge-neutral by itself, but is unstable since any background doping (intentional or unintentional) converts the system into an extrinsic or doped semiconductor with either free electrons in the conduction band or free holes in the valence band (depending on the nature of background doping). Finite temperature also creates free carriers in the system, albeit maintaining its overall charge neutrality since equal numbers of thermally excited electrons and holes are created.  It is therefore more appropriate to think of 3D Dirac materials as gapless semiconductors rather than semimetals since the stable system is a doped or extrinsic semiconductor (with linear chiral energy bands) with free carriers.  In any semiconductor, including 3D Dirac systems, the most important physical property is the nature of charge transport which is easy to measure experimentally as a function of temperature, carrier density, and other variables (e.g., external magnetic field, applied pressure, etc.).  Drude or ohmic dc transport as a function of carrier density and temperature provides detailed information about the elementary excitations and scattering processes in the system, and is thus the very first property typically studied in the literature.  Indeed, transport studies of 3D metals and semiconductors as well as two dimensional electron gas systems and graphene have been among the most active research areas in physics and applied physics over the last sixty years (since the beginning of the 1950s).

Recent work on graphene, the quintessential 2D Dirac material, shows the great importance of transport properties in the context of Dirac materials. \cite{geim2007,neto,peres,dassarma2011}   The nature of transport in graphene has been actively debated \cite{jang2008} over the last seven years, and the nature of the graphene Dirac point has been elucidated through experimental and theoretical works studying density and temperature dependent electrical conductivity.\cite{geim2007,neto,peres,dassarma2011} More recently, transport properties of 2D surface states on 3D topological insulators have been studied in the context of understanding their fundamental properties.\cite{ti}   It is clear that understanding the transport properties of an electronic material is essentially the first step in developing a complete theory for the material itself.  Motivated by these considerations, we undertake in this work the development of a theory for the density and temperature dependent Drude-Boltzmann transport properties of 3D Dirac materials using charged impurity and acoustic phonon scattering processes as the main resistive mechanisms.  Our theory includes the full effect of finite temperature screening due to the chiral linear band structure of the Dirac system.
The corresponding theoretical and experimental transport work on graphene (i.e., the most common 2D Dirac material) is extensively reviewed in Refs. \onlinecite{peres,dassarma2011}.

In the current work,
we explore the transport properties of 3D Dirac systems theoretically,
obtaining the dc conductivity as a function of both carrier density
($n$) and temperature ($T$) considering both the undoped (intrinsic
semimetal with the chemical potential precisely at the Dirac point)
and the doped (extrinsic semiconductor with the chemical potential in
the conduction band at a Fermi energy $E_F$ determined by the doping
density). 
There have been several recent publications discussing transport properties of these 3D chiral materials, \cite{hosur2012,sbierski2012,skinner, wan2001,biswas2014}  but the investigation including full temperature and density dependent screening effects on transport of 3D Dirac semimetals, as we consider in the current work,
has not been considered in the literature for 3D Dirac materials.
In addition, we provide a comparison among several different momentum relaxation processes (e.g. disorder and phonon scattering, long-range and short-range disorder, transport and quantum relaxation, etc.), which
has also not been discussed. In this paper we calculate transport properties incorporating the density and temperature dependent screening
of random background quenched charged impurities, typically the most common disorder in semiconductors.  We also include scattering by zero-range white-noise disorder arising, for example, from point defects.
We also consider the temperature dependent conductivity limited by the acoustic phonon scattering in 3D Dirac materials. Similar transport theory results coupled with corresponding
experimental conductivity measurements \cite{neto, peres,
  dassarma2011, hwang2007, tan2007,
  adam2007, chen2008, jang2008} led to considerable fundamental
understanding and progress in graphene physics as well as for surface transport in 3D topological insulator materials\cite{ti}, and the expectation is
that the same should happen in 3D Dirac materials when appropriate
transport measurements are carried out and compared with the
theoretical results presented in the current work.
In particular and as emphasized above, a thorough theoretical understanding of carrier transport properties is almost inevitably the first basic requirement in developing the physics and materials science of any new doped semiconductors.  This is the context of our current theoretical work exploring the transport properties of doped 3D Dirac materials which are a particular type of semiconductors with properties distinct from ordinary parabolic semiconductors.
Our work should be useful in understanding and interpreting experimental transport data in 3D Dirac materials as they become available in the future.

The theoretical approach we develop is physically motivated and based
on the firm (and remarkably successful) conceptual framework of carrier
transport theories in bulk semiconductors and semimetals as well as in
2D semiconductor systems\cite{andormp} and
graphene\cite{dassarma2011}. We use the semiclassical 
Boltzmann transport theory within the relaxation time approximation
for solving the Boltzmann integral equation. We take into account
resistive scattering by long-range and short-range impurity
scattering as well as phonon scattering. The long-range disorder arises from random quenched
charged impurities in the environment, which are invariably present
either as dopant ions to produce free carriers in the system or as
unintentional background impurities in the host material. The short-range disorder arises
from atomic point defects which may be present in the system. 
The Coulomb disorder
arising from the quenched charged impurities is screened by free Dirac carriers themselves which we treat within the finite temperature random phase
approximation (RPA).
We note that the resistivity due to long-range unscreened Coulomb disorder typically diverges in 3D electronic systems, and therefore, appropriate screening is an essential element of the theory in order to get meaningful finite (rather than vanishing) conductivity in the system.
Phonons are inevitable sources of scattering and can dominate transport at high temperatures. We therefore consider the phonon limited conductivity incorporating the acoustic phonon scattering in the 3D Dirac semimetals which could be of importance at higher temperatures
whereas the low-temperature conductivity should always be dominated by disorder.  Our theory includes screening, disorder and phonons on an equal footing for the calculation of the transport properties of 3D Dirac systems within the Boltzmann-RPA transport theory.

In general, the conductivity is related to the transport relaxation time $\tau_t$ in disordered metals, i.e., $\sigma \propto \tau_t$. There is another distinct relaxation time in a disordered system that is  independent of $\tau_t$, which is the single particle (or quantum) relaxation time $\tau_s$ defining the quantum level broadening. \cite{mahan,gant,dassarma1985,hwang_scattering} Although $\tau_t$ and $\tau_s$ both arise from impurity scattering in the metallic
regime, they are distinct from each other with no direct analytical relationship connecting them except
for the simple model of completely isotropic s-wave zero-range impurity scattering where they become identical.
It is well known\cite{gant} that in ordinary 3D metals $\tau_t \approx \tau_s$ because of the strong screening effect (i.e., the effective impurity-electron interaction potential is short ranged).
However, 3D Dirac semimetals are qualitatively different even for this short-ranged white noise disorder due to their chiral symmetry, which suppresses backward (i.e., a scattering induced wave vector change by $2k_F$) scattering. 
In addition, screening is weak in Dirac systems due to the small density of states associated with the linear band dispersion, and hence long-range Coulomb disorder could introduce differences between transport and  quantum relaxation times, drawing a sharp contrast with 3D semiconductors (and metals) where $\tau_t \sim \tau_s$ typically holds even for Coulomb disorder due to strong carrier screening. 
In this paper, we theoretically study $\tat$ and $\tas$ in 3D Dirac semimetals due to long-range impurity scattering, finding interesting behavior in the ratio $\tas/\tas$ as a function of the effective coupling constant. We compare the ratio $\tau_t/\tau_s$ of 3D Dirac semimetals with the corresponding situation in 2D graphene \cite{hwang_scattering} which is a typical 2D Dirac material, finding significant differences.
As we will show later in this paper, due to the chiral suppression of backward scattering, the zero-range white-noise s-wave isotropic disorder potential leads to $\tau_t/\tau_s = 3/2$ in 3D Dirac semimetals, in contrast to the $\tau_t = \tau_s$ condition in ordinary metals and semiconductors for short-range disorder. 
For 3D Dirac semimetals with the Coulomb disorder being strongly screened by the RPA screening  we have $\tau_t/\tau_s \geq 3/2$ depending on the ultraviolet cutoff of the linear band dispersion, even though for the corresponding Thomas-Fermi (TF) screening $\tau_s/\tau_s \rightarrow 3/2$ in the strong screening limit.
This interesting (and nontrivial) difference between RPA and Thomas-Fermi screening in Dirac systems arises from the subtle property of the chiral nature of the free carriers which suppresses the $2k_F$ back-scattering.

The paper is organized as follows. In Sec. II, the Boltzmann transport theory is presented to calculate the temperature and density dependent conductivity of 3D Dirac semimetals in the presence of Coulomb disorder (i.e. random charged impurities). We also present screening (by the carriers themselves) within the RPA theory. 
Section III presents the results for the transport scattering time $\tau_t$ and the single particle relaxation time $\tau_s$. In Sec. IV, we present the acoustic phonon scattering limited resistivity of 3D Dirac semimetals. In Section V we discuss the results and predict experimental implications.

\section{screening and conductivity}

\subsection{RPA and TF screening}

Before we discuss the temperature- and density-dependent conductivity we first consider temperature-
and density-dependent screening (or dielectric function) since the Coulomb disorder
arising from the quenched charged impurities is screened by free Dirac carriers themselves
(i.e., the bare Coulomb disorder $V$ becomes screened Coulomb disorder $V/\epsilon$ where $\epsilon$ is the dielectric function).
Within the RPA the
screening function (static 
dielectric function) is given by 
\begin{equation}
\epsilon(q,T) = 1- v(q) \Pi(q,T),
\label{dielec}
\end{equation}
where $v(q) = 4\pi e^2/\kappa q^2$ is the 3D Fourier transform of the
Coulomb potential $e^2/\kappa r$ and $\Pi(q,T)$ is the static polarizability
which is given by 
\begin{equation}
\Pi(q,T) = \frac{D_F}{6}  \frac{q^2}{k_F^2}  \ln  \frac  {k_c}{q} +
\Pi^{(+)}(q,T) + \Pi^{(-)}(q,T), 
\label{pit0}
\end{equation}
where $D_F = g E_F^2/[2\pi^2(\hbar v_F)^3]$ is the density of states
at the Fermi level, $k_c$ is the ultraviolet cutoff arising from the
linear band dispersion ($k_c \sim a^{-1}$ where $a$ is the lattice constant),
showing that there is an ultraviolet renormalization of effective charge in 3D Dirac systems in contrast to the 2D Dirac system where the charge remains unrenormalized
(i.e. the corresponding 2D Dirac polarizability function does not have any ultraviolet cut off correction \cite{dassarma2011}),
and $\Pi^{(\pm)}$ are given by 
\begin{equation}
\Pi^{(\pm)}(q,T)= \frac{D_F}{k_F^2} \int_0^{\infty} dk f_k^{(\pm)} \left [ k -
  \frac{q^2-4k^2}{4q} \ln \left | \frac {q+2k}{q-2k} \right | \right
], 
\label{pit}
\end{equation}
where $f_k^{(\pm)} = 1/[1+\exp[(\varepsilon_k \mp \mu)/k_BT]]$, where
$\varepsilon_k = \hbar v_F k$ and $\mu$ is the finite temperature
chemical potential which is  determined by the conservation of the
total electron density.  Note that only cutoff $k_c$ dependance in the screening function arises from the first term in Eq.~(\ref{pit0}) which is independent of the temperature 
-- thus, temperature ($T$) and ultraviolet cut off ($k_c$) do not mix at all in the theory.
Thus,  it is expected that the ultraviolet cutoff is irrelevant for the temperature dependent conductivity within RPA
although it will manifest itself as a weak logarithmic effect in the zero-temperature resistivity arising from the scattering by the screened Coulomb disorder.  We will suppress this ultraviolet logarithmic dependence in most of our discussions below.
At zero temperature $\mu(T) = E_F$, and  $\Pi^{(-)}(q) = 0$ and 
\begin{eqnarray}
{\Pi}^{(+)} (q) = \frac{2D_F}{3} \left [ 1  +  \frac{k_F}{2q} \left
  (1-\frac{3}{4}\frac{q^2}{k_F^2} \right ) \ln \left |\frac{2k_F +
    q}{2k_F-q} \right | \right . \nonumber \\ 
 \left . -   \frac{q^2}{8k_F^2} \ln \left | \frac{4k_F^2 - q^2}{q^2}
 \right | \right ]. \;\;\;  
\end{eqnarray}
In the low temperature limit ($T<T_F$) the asymptotic forms of the
polarizability become at $q=0$ 
\begin{equation}
\tilde{\Pi}(0,T) = 1 - \frac{\pi^2}{3} \left ( \frac{T}{T_F} \right ) ^2 ,
\end{equation}
and at $q=2k_F$
\begin{eqnarray}
\tilde{\Pi}(2k_F,T) & =  & \tilde{\Pi}(2k_F,0)  \nonumber \\
& - & \frac{\pi^2}{6} \left [ 1 + \ln \frac{\pi^2}{12} + \ln
  \frac{T}{T_F} \right ] \left ( \frac{T}{T_F} \right )^2 , 
\end{eqnarray}
where $\tilde{\Pi} = \Pi /D_F$,
and $\tilde{\Pi}(2k_F,0) = \frac{2}{3} [1 + \ln (k_c/2k_F)]$
At high temperatures ($T>T_F$) we have
\begin{equation}
\tilde{\Pi}(0,T) = \frac{\pi^2}{3} \left ( \frac{T}{T_F} \right )^2 +
\frac{1}{\pi^4} \left ( \frac{T_F}{T} \right )^4. 
\end{equation}

Before considering the full RPA screening effect we first consider TF
screening which is defined as the $q=0$ limit of RPA. The temperature
dependent TF wave vector can be calculated 
at low temperatures ($T/T_F < 1$) 
\begin{equation}
q_{TF}(T) = q_{TF}(0) \left [1-\frac{\pi^2}{6} \left (
  \frac{T}{T_F} \right )^2 \right ], 
\label{qtflow}
\end{equation}
and at high temperatures ($T/T_F > 1$)
\begin{equation}
q_{TF}(T) = q_{TF}(0)  \frac{\pi}{\sqrt{3}} \left ( \frac{T}{T_F}
  \right ) \left [ 1 + \frac{3}{2\pi^6} \left ( \frac{T_F}{T} \right
    )^6 \right ], 
\label{qtfhigh}
\end{equation}
where $q_{TF}(0) = \sqrt{2 g \alpha/\pi} k_F$ is the zero temperature
TF wave vector.
We note that the TF screening wave vector is simply related to the density of states at the Fermi energy through the relationship: $q_{TF}(0)= (4\pi e^2 D_F/\kappa)^{1/2}$.

\begin{figure}[t]
	\centering
	\includegraphics[width=1.\columnwidth]{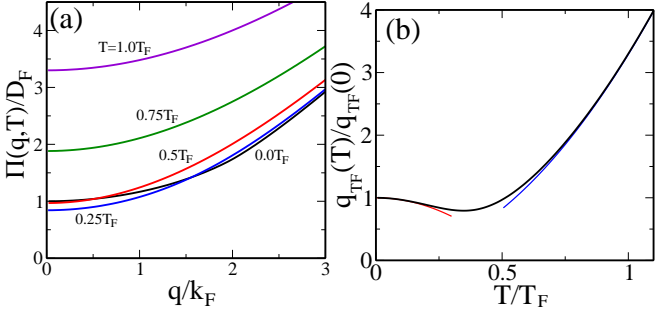}	
	\caption{
(a) Temperature-dependent polarizability $\Pi(q,T)$ of 3D Dirac semimetals as a function of wave vector for different temperatures and (b) temperature dependent TF wave vector $q_{TF}(T)$ as a function of temperature. Here $D_F$ is the density of states of the 3D Dirac semimetals and $q_{TF}(0)$ is the zero temperature TF screening wave vector. 
The low and high temperature behaviors given in  Eq.~(\ref{qtflow}) and Eq.~(\ref{qtfhigh}), respectively, are also shown in (b). 
The TF screening wave vector $q_{TF}(0)$ is related to the density of states at the Fermi energy through the relationship: $q_{TF}(0)= (4\pi e^2 D_F/\kappa)^{1/2}$, and by definition $q_{TF}^2=\Pi (q=0) 4\pi e^2/\kappa$.
}
\label{fig1}
\end{figure}

In Fig.~\ref{fig1}, we show our calculated 3D Dirac RPA (sometimes called Lindhard) screening function for various temperatures.  We note that there are significant differences with the corresponding 2D \cite{2Dscreening} and 3D \cite{mahan,lindhart}  Lindhard screening function
for ordinary metals (i.e., parabolic non-chiral 3D systems), but many similarities with 2D Dirac systems (e.g., graphene) \cite{screening}.
As shown in Fig.~\ref{fig1} the overall screening function for $T \agt 0.5T_F$ increases monotonically with temperature because of the thermal excitation of electrons from the valence band to the conduction band. This behavior is also found in graphene \cite{screening} and is generic for gapless semimetals.
The overall temperature dependence of 3D Lindhard screening function decreases with temperature for $q \alt 2k_F$, but increases with temperature for $q \gg 2k_F$ because the thermally excited electrons only in the conduction band weakens the long wavelength screening.
The TF screening wave vector also shows different behavior from the ordinary metals where the TF wave vector decreases monotonically with temperature. However, the TF wave vector in 3D Dirac semimetals increases linearly with temperature at high temperatures ($T \gg T_F$), but decreases quadratically at low temperatures ($T < T_F$). The TF screening in graphene also shows a similar behavior \cite{screening}.
We note that the scaled screening function ($\Pi(q/k_F,T/T_F)/N_F$) in terms of $q/k_F$ and $T/T_F$ is very weakly dependent on the carrier density, and in fact the density dependent term of scaled $\Pi$ arises only from the ultraviolet cutoff in Eq.~(\ref{pit0}), i.e., $\ln(k_c/k_F)$.

\subsection{Boltzmann Conductivity}

For a system with a chiral linear energy dispersion, $E_k = \hbar v_F |\vk|$,
the energy dependent transport scattering time within the Born
approximation is given by 
\begin{equation}
\frac{1}{\tau_t} = \frac{2\pi n_i}{\hbar}\int \frac{d^3\vk'}{(2\pi)^3}
\left |\langle V(\vk,\vk') \rangle \right |^2
\frac{(1-\cos^2\theta)}{2} \delta(E_k-E_{k'}), 
\label{tauin}
\end{equation}
where $n_i$ is the impurity density, $\theta$ is the scattering angle
between $\vk$ and $\vk'$, and 
$\langle V(\vk,\vk') \rangle$ is the scattering amplitude. For
screened Coulomb disorder we have
\begin{equation}
\langle V(\vk,\vk') \rangle = \frac{4\pi e^2}{\kappa} \frac{1}{q^2 +
  q_s^2},
\end{equation}
where $\kappa$ is the background lattice dielectric constant of the material,
$q = |\vk-\vk'|$, and $q_s$ is the screening wave vector. 
For $q_s=0$, we get the usual unscreened Coulomb potential in the wave vector space going as $1/q^2$ indicating the long-range $1/r$ behavior of the Coulomb interaction.
We can rewrite Eq.~(\ref{tauin}) as
\begin{equation}
\frac{1}{\tau_t} = \frac{n_i}{2\pi} \frac{k^2}{\hbar^2 v_F} 
  \int_0^{\pi}d\theta |V(q)|^2\sin\theta(1-\cos^2\theta)/2,
\end{equation}
where $q = 2k \sin(\theta/2)$.

In the long wavelength Thomas-Fermi approximation, the screening wave vector $q_s$ is defined by the density of states at the Fermi energy.
Considering Thomas-Fermi (TF) screening we have the scattering time for
long range Coulomb disorder at $T=0$
\begin{equation}
\frac{1}{\tau_t} = 4\pi n_i \alpha^2 \frac{v_F}{k_F^2} I_t(q_0)
\label{tauin2}
\end{equation}
where $\alpha = e^2/\kappa \hbar v_F$ is the effective coupling (i.e.,
fine structure) constant and
$q_0 = q_{TF}/2k_F$ with 
$q_{TF} = \sqrt{12 \pi e^2 n/E_F} = \sqrt{2g\alpha/\pi} k_F$ being the
TF screening wave vector ($g=g_sg_v$ is the total degeneracy with
$g_s$ and $g_v$ being the spin and valley degeneracy, respectively).  
In Eq.~(\ref{tauin2}) $I_t(q_0)$ is given by
\begin{equation}
I_t(q_0)  =  \left (q_0^2 + \frac{1}{2} \right ) \log \left ( 1 + \frac{1}{q_0^2}
\right ) -1. 
\end{equation}
The asymptotic behaviors of $I(q_0)$ are given by
\begin{eqnarray}
I_t(q_0) & \sim & -1 - \log(q_0) + q_0^2/2 - 2q_0^2 \log(q_0), \;\; {\rm for}
\; q_0 \ll 1  \nonumber \\
I_t(q_0) &\sim & \frac{1}{12q_0^4} \left ( 1 - \frac{1}{q_0^2} \right ),
\;\; {\rm for} \; q_0 \gg 1. 
\end{eqnarray}
Note that $q_0 = q_{TF}/2k_F = \sqrt{g\alpha/2\pi}$ is a function of
the coupling constant $\alpha$ and is independent of density.  
Since $n = g k_F^3 /6\pi^2$, the density dependent scattering time is given by 
\begin{equation}
\frac{1}{\tau_t} \propto \frac{n_i}{n^{2/3}}.
\end{equation}

For the Dirac materials the zero temperature conductivity $\sigma$ can
then be expressed in the TF approximation as
\begin{equation}
\sigma = e^2 \frac{v_F^2}{3} D_F \tau_t = \frac{e^2}{h} \frac{ g
  v_F}{3\pi} k_F^2 \tau_t, 
\label{sigma}
\end{equation}
where $D_F = g E_F^2/[2\pi^2 (\hbar v_F)^3]$ is the density of states at
the Fermi level.
With Eq.~(\ref{tauin2}) we have 
\begin{equation}
\sigma =\frac{e^2}{h}\frac{g}{12\pi^2}\frac{k_F^4}{n_i\alpha^2}
\frac{1}{I_t(q_0)}.
\label{sigmat0}
\end{equation}
Thus, for $q_0 \ll 1$
\begin{equation}
\sigma = \frac{e^2}{h} \left ( \frac{3\pi^2}{4 g} \right )^{1/3}
\frac{n^{4/3}}{\alpha^2 n_i}\frac{1}{\log(1/q_0) -1},
\label{sigmalow}
\end{equation}
and for $q_0 \gg 1$
\begin{equation}
\sigma = \frac{e^2}{h}\frac{3g}{\pi}\left ( \frac{3g^2}{4\pi} \right
)^{1/3} \frac{n^{4/3}}{n_i}.
\label{sigmahi}
\end{equation}
For $q_0 = \sqrt{g\alpha/2\pi} \gg 1$,  the conductivity is
independent of the coupling constant $\alpha$.
This result also corresponds to the complete screening of the Coulomb
disorder, i.e.,
\begin{equation}
\langle V(\vk,\vk') \rangle = \frac{4\pi e^2}{\kappa} \frac{1}{q_s^2}.
\end{equation}
With the completely  screened Coulomb disorder the scattering time becomes
\begin{equation}
\frac{1}{\tau_t} = \frac{4\pi^3 n_i v_F}{3g^2 k_F^2}.
\end{equation}
Substituting this result into Eq.~(\ref{sigma}) we have the same
result as Eq.~(\ref{sigmahi}).

\begin{figure}[t]
	\centering
	\includegraphics[width=.8\columnwidth]{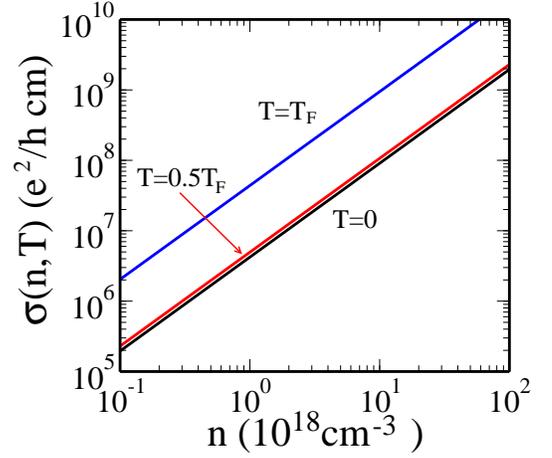}
	\caption{
The density dependent conductivity calculated with the full RPA
screening function for various temperatures. 
Here $\alpha = 1.2$ and a fixed impurity density
$n_i = 10^{18} cm^{-3}$ are used. 
}
\label{fig0}
\end{figure}

For the short range 
(actually, zero-range, as appropriate for point defects and neutral impurities)
disorder with 
$\langle V(\vk,\vk') \rangle = V_0$ we have
\begin{equation}
\frac{1}{\tau_t} = \frac{n_d V_0^2 k_F^2}{3\pi \hbar^2 v_F},
\label{tauins}
\end{equation}
and the conductivity becomes
\begin{equation}
\sigma = \frac{e^2}{h} \frac{g (\hbar v_F)^2}{n_d V_0^2}.
\end{equation}
For the short ranged disorder the conductivity is independent of the
carrier density whereas the long-range disorder gives $\sigma \propto
n^{4/3}$ (see Fig.~\ref{fig0}).
A peculiar feature of the 3D strong screening limit is that the conductivity becomes independent of the effective fine structure constant, which is qualitatively different from the corresponding 2D graphene case.
In Table I we summarize our calculated exponent ($\beta$) of the density dependent conductivity ($\sigma \sim n^{\beta}$) for the various types of disorder present in the environment of the 3D Dirac materials comparing with that of 2D graphene and 3D parabolic systems \cite{dassarma_scaling}.
As is obvious from Table 1 (and the theory above), 3D Dirac systems show qualitatively different density scaling of the electrical conductivity compared with either ordinary (parabolic) 3D metals/semiconductors or 2D graphene.

\begin{center}
\begin{table*}[t]
\caption{\label{tab:table1} 
By considering the various types of disorders present in the environment of 3D Dirac materials, the asymptotic density scaling exponent ($\beta$) for the conductivity ($\sigma \sim n^{\beta}$ in 3D Dirac systems) is compared with the corresponding exponents of other systems \cite{dassarma_scaling}.
The two values of the exponents for the 2D Dirac system in the table indicate the impurities being distributed inside (far outside) the 2D system.  See Ref. \onlinecite{dassarma_scaling} for details on the results given in the second and third columns whereas the results of the first column (for 3D Dirac systems) are obtained in the current work.  The weak- or strong-screening limits imply $q_0=q_{TF}/2k_F \ll 1$ or $\gg1$ -- see the text for details.
}
\hspace{0.1\hsize}
\begin{ruledtabular}
\begin{tabular}{|l|c|c|c|}
Disorder  type &  3D Dirac system &   2D Dirac system   &  Ordinary 3D   \\ \hline 
Unscreened Coulomb &  Log-divergent &  Log-divergent (0) &  Log-divergent  \\ 
Screened Coulomb (weak) & 4/3 & 1/2  (0) & 1  \\ 
Screened Coulomb (strong) & 4/3  & 1/2 (0) & 1/3 \\ 
Zero range & 0 & -1 (-1) & -1/3  \\
\end{tabular}
\end{ruledtabular}
\hspace{0.1\hsize}
\end{table*}
\end{center}

In the Boltzmann transport theory the temperature dependence of the
conductivity arises from two separate contributions
\cite{dassarma2011,andormp}, namely the temperature 
dependent screening and energy averaging of the energy dependent
scattering time. 
Since the energy dependence of the scattering time for the long range
Coulomb disorder is given by
$\tau_t(E) \propto E^2$ we have the energy averaged scattering time at
low temperatures
\begin{equation}
\langle \tau_t(T) \rangle = \tau_t(0) \left [ 1 + \frac{2\pi^2}{3} \left
  (\frac{T}{T_F} \right )^2 \right ],
\label{taulow}
\end{equation}
and at high temperatures 
\begin{equation}
\langle \tau_t(T) \rangle = \tau(0) \frac {7 \pi^4}{30} \left (
\frac{T}{T_F} \right )^4,
\label{tauhigh}
\end{equation}
where $\tau_t(0)$ is the zero temperature scattering time given in
Eq.~(\ref{tauin2}). 
Thus, combining the temperature dependent scattering time with the
energy averaged terms we have the total
temperature dependent conductivity.
For the long range disorder the temperature dependent screening wave
vector gives rise to an additional temperature dependence in the
conductivity. Thus, the $\tau_t(0)$ in Eqs.~(\ref{taulow}) and
(\ref{tauhigh}) can be considered to be an effective temperature dependent scattering
time arising from the temperature dependent screening effect as derived above.

\begin{figure}
	\centering
	\includegraphics[width=.8\columnwidth]{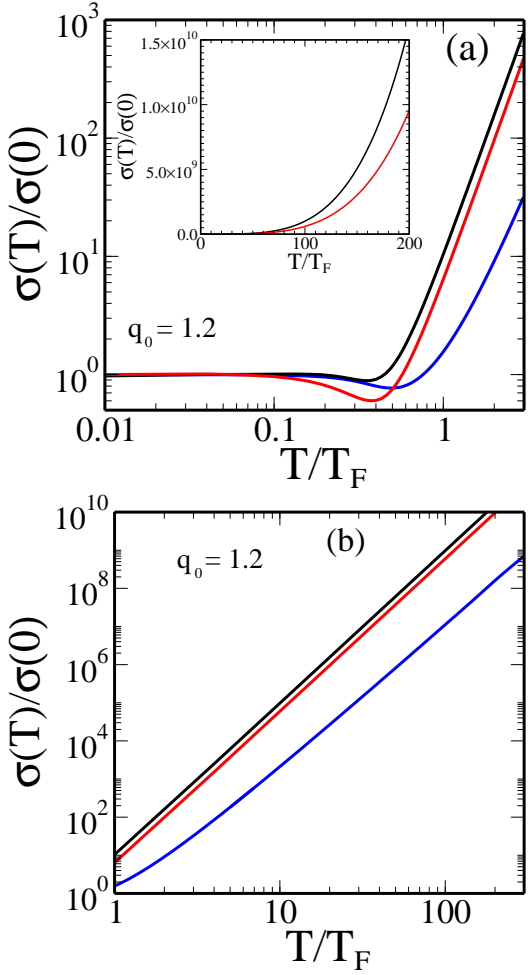}
	\includegraphics[width=.8\columnwidth]{fig_f2b.eps}
	\caption{
The scaled conductivity $\sigma(T)/\sigma(T=0)$ for a strong screening
parameter $q_0 = q_{TF}/2k_F = 1.2$ (a) at low temperatures and (b) at
high temperatures. Inset in (a) shows the same figures as (a) and (b)
in the linear scale. The black line indicates the conductivity
calculated with the full RPA screening. The red (blue) line
indicates the conductivity calculated with temperature dependent
(independent) TF.
}
\label{fig3}
\end{figure}

\begin{figure}
	\centering
	\includegraphics[width=.8\columnwidth]{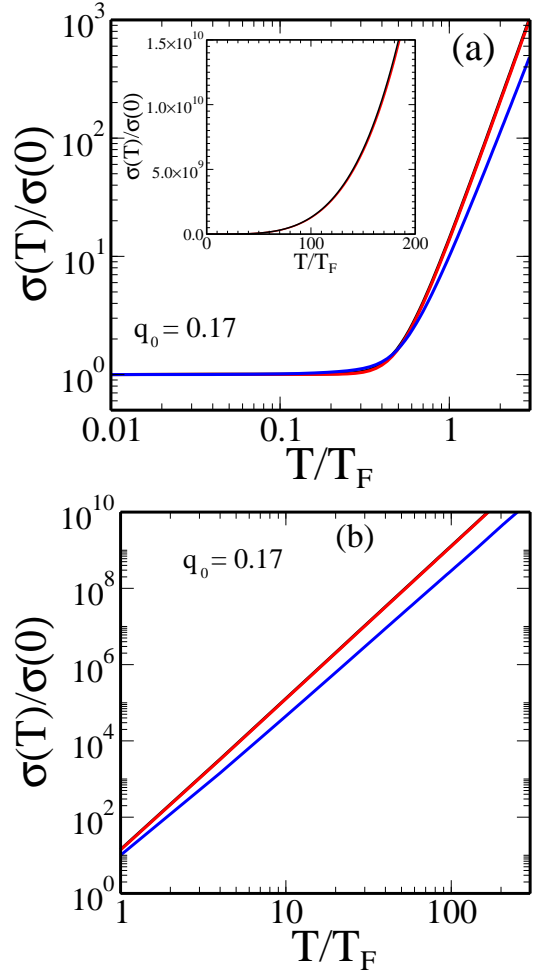}
	\includegraphics[width=.8\columnwidth]{fig_f3b}
	\caption{
The same as Fig.~\ref{fig3} for a weak screening
parameter $q_0 = q_{TF}/2k_F = 0.17$.
}
\label{fig4}
\end{figure}

We can now obtain the overall temperature dependence of the
conductivity by combining the energy averaging and temperature
dependent screening.
For $q_0<1$  Eqs.~(\ref{sigmalow}) and (\ref{taulow}) give the low
temperature conductivity as
\begin{equation}
\sigma(T) = \sigma(0)\left[ 1+\frac{\pi^2}{3} \left ( 2 + \frac{1}{ 1
    + \ln q_0} \right ) \frac{T^2}{T_F^2} \right ],  
\end{equation}  
where $\sigma(0)$ is defined by Eq.~(\ref{sigmat0}) which gives the density-dependent conductivity at zero-temperature, 
and Eqs.~(\ref{sigmalow}) and (\ref{tauhigh}) give the high
temperature conductivity for $q_0 T/T_F <1$ 
\begin{equation}
\sigma(T) = \sigma(0) \frac{7\pi^4}{30} \left ( \frac{T}{T_F} \right
)^4 \left [ \frac{1+ \ln q_0}{1+\ln (\pi q_0/\sqrt{3}) +  \ln (T/T_F)
  } \right ], 
\end{equation} 
and for $q_0T/T_F > 1$ we have 
\begin{equation}
\sigma(T) = \sigma(0) \frac{7 \pi^8}{90} \left ( \frac{T}{T_F} \right )^8.
\end{equation}
Thus, for $q_0< 1$ there is a crossover in the temperature dependence
from an exponent 4 to 8 in the high temperature conductivity. 
For $q_0 > 1$ we have the low temperature  conductivity
\begin{equation}
\sigma(T) = \sigma(0) \left [ 1 + O(T/T_F)^4 \right ]
\label{siglowqhi}
\end{equation}
and the high temperature conductivity
\begin{equation}
\sigma(T) = \sigma(0) \frac{7 \pi^8}{90} \left ( \frac{T}{T_F} \right )^8.
\label{sighiqhi}
\end{equation}
Due to the cancelation between the screening effect and the energy
averaging, the low temperature conductivity is almost temperature
independent.  
The $T^8$ dependence of conductivity in the high $q_0T/T_F$ limit is
completely artificial, arising from the independent considerations of
the screening and energy averaging. As shown in Figs.~\ref{fig3} and
\ref{fig4}  the energy 
dependence of the screening gives $T^4$ behavior in the high
temperature limit even for the TF screening.

For the short range disorder we have the energy averaged scattering
time at low temperatures
\begin{equation}
\langle \tau_t(T) \rangle = \tau_t(0) \left [ 1 - e^{-T_F/T}
  \right ],
\end{equation}
and at high temperatures
\begin{equation}
\langle \tau_t(T) \rangle = \tau_t(0) \left [ \frac{1}{2} +
  \frac{1}{4\pi^2} \left ( \frac{T_F}{T} \right )^3 \right ],
\end{equation}
where $\tau_t(0)$ is given in Eq.~(\ref{tauins}).
For the short range disorder the temperature dependent conductivity
comes from the energy averaging of the scattering time because there
is no screening effect.

In Figs.~\ref{fig3} and \ref{fig4} we show our full numerically
calculated RPA screened (i.e., full wave vector, temperature, and
density dependent static screening) transport results for 3D Dirac
systems, emphasizing that the asymptotic low- and high-temperature
results agree well with our analytical theories.
A particularly noteworthy and unexpected feature is that, except for a
small window of temperature around $T/T_F \sim 0.5$ where the conductivity
decreases with increasing temperature, the impurity-induced transport
behavior in 3D Dirac materials is `insulating' at all temperatures
with the conductivity increasing with increasing temperature.
Since screening always decreases with increasing temperature, our results indicate that the temperature dependence of the electrical conductivity of 3D Dirac systems is mostly dominated by the energy averaging effects, leading mostly to a decreasing conductivity with decreasing temperature as long as phonon effects and any electron interaction-induced umklapp scattering processes can be ignored.  Phonons by themselves should lead to very weak temperature dependence at low temperatures (the so-called Bloch-Gr\"{u}neisen regime) as studied later in this work whereas the combination of umklapp scattering and electron-electron interaction should produce a $T^2$ temperature dependence decreasing the conductivity with increasing temperature.
We note that the temperature dependent conductivity within RPA is unaffected by
the ultraviolet cutoff $k_c$ because $k_c$ only enters in the first term of the RPA screening function, Eq.~(\ref{pit0}), and this term is independent of temperature.


\section{transport scattering time and single particle relaxation time}

In previous sections we calculated the transport scattering time $\tau_t$ which determines the conductivity. In this section we consider the quantum lifetime or the single-particle relaxation time, $\tau_s$, which determines the quantum level broadening $\Gamma \equiv \hbar/2\tau_s$ of the momentum eigenstates. \cite{mahan, gant}
The single particle relaxation time
$\tau_s$ can be calculated from the electron self energy of the  
coupled electron-impurity system. It is related to the imaginary part
of the single particle self-energy function by
\begin{equation}
\frac{1}{\tau_s} =\frac{2}{\hbar} {\rm Im} \Sigma(k_F,E_F),
\end{equation} 
with the single particle quantum (impurity induced) level broadening
$\Gamma_s = {\rm Im}\Sigma(k_F,E_F)$, i.e. $\tas = \hbar/2\Gamma_s$.
In the leading order disorder approximation,
we obtain the single particle relaxation time $\tau_s$
\begin{equation}
\frac{1}{\tau_s} = \frac{2\pi n_i}{\hbar}\int \frac{d^3\vk'}{(2\pi)^3}
\left |\langle V(\vk,\vk') \rangle \right |^2
\frac{(1+\cos\theta)}{2} \delta(E_k-E_{k'}).
\label{scat_s}
\end{equation}
By comparing Eq.~(\ref{tauin}) with Eq.~(\ref{scat_s}), we find that the only difference between the scattering time $\tat$ and the
single-particle relaxation time $\tas$ is the weighting factor for the backscattering ($1-\cos 
\theta$) in the transport scattering time which arises from the vertex correction which must be present in the conductivity.
Thus, without the chiral factor for the wave function overlap ($1+\cos \theta$), the difference between $\tau_t$ and $\tau_s$ arises from the effect
of the wave vector dependent impurity potential $V(\vk)$ which
distinguishes between single particle relaxation $\tau_s^{-1}$ which is affected equally by scattering in all directions and transport relaxation $\tau_t^{-1}$ which
is unaffected by forward scattering (small-angle
scattering).  Note that although the transport scattering time, $\tau_t$,  and the single particle scattering time $\tau_s$ both arise from impurity scattering, they are in general distinct and unique with no direct analytical relationship connecting them (except when the disorder potential is isotropic or zero-ranged).
In addition to the weighting factor for the backscattering, there is another weighing factor for the chiral effect in 3D Dirac semimetals, ($1+\cos\theta$), which suppresses the large angle scattering equally for both $\tau_t$ and $\tau_s$. 

Considering Thomas-Fermi (TF) screening we have the scattering time for
long range Coulomb disorder at $T=0$
\begin{equation}
\frac{1}{\tau_s} = 4\pi n_i \alpha^2 \frac{v_F}{k_F^2} I_s(q_0)
\label{tauins2}
\end{equation}
where $\alpha = e^2/\kappa \hbar v_F$ is the effective coupling (i.e.,
fine structure) constant and
$q_0 = q_{TF}/2k_F$ with 
$q_{TF} = \sqrt{12 \pi e^2 n/E_F} = \sqrt{2g\alpha/\pi} k_F$ being the
TF screening wave vector ($g=g_sg_v$ is the total degeneracy with
$g_s$ and $g_v$ being the spin and valley degeneracy, respectively).  
In Eq.~(\ref{tauins2}) $I_s(q_0)$ is given by
\begin{equation}
I_s(q_0)  =  \frac{1}{4} \left [ \frac{1}{q_0^2} - \log \left ( 1 + \frac{1}{q_0^2} \right )
\right ]. 
\end{equation}
The asymptotic behaviors of $I_s(q_0)$ are given by
\begin{eqnarray}
I_s(q_0) & \sim & \frac{1}{4q_0^2} \left [ 1 + 2 q_0^2 \log q_0 - q_0^2 \right ], \;\; {\rm for}
\; q_0 \ll 1  \nonumber \\
I_s(q_0) &\sim & \frac{1}{8q_0^4} \left ( 1 - \frac{1}{3q_0^2} \right ),
\;\; {\rm for} \; q_0 \gg 1. 
\end{eqnarray}
Note that $q_0 = q_{TF}/2k_F = \sqrt{g\alpha/2\pi}$ is a function of
the coupling constant $\alpha$ (in 3D Dirac semimetals the dimensionless interaction parameter is 
$\alpha$) and is independent of density.  
Thus, the density dependent scattering time has the same form as $\tau_t$
\begin{equation}
\frac{1}{\tau_s} \propto \frac{n_i}{n^{2/3}}.
\end{equation}

From Eqs.~(\ref{tauin2}) and (\ref{tauins2}) we can obtain the ratio of the transport scattering time to the single particle relaxation time, $\tau_t/\tau_s$. Note that the ratio is only a function of $q_0$ (or $\alpha$) and does not depend on the carrier density. We have the limiting form of
$\tau_t/\tau_s$, in the small and large $q_0$ regimes
\begin{eqnarray}
\frac{\tau_t}{\tau_s} & \approx & \frac{3}{2} \left ( 1 + \frac{1}{3q_0^2} \right ), 
\;\; {\rm for} \; q_0 \gg 1 \nonumber \\
\frac{\tau_t}{\tau_s} & \approx & \frac{1}{2 q_0^2 (-1-\log q_0)} \left [ 1+ 2 q_0^2 \log q_0 \right ]
\;\; {\rm for} \; q_0 \ll 1.
\end{eqnarray}
For large $q_0$, the ratio approaches 3/2, which indicates that scattering is not isotropic even though the screening is strong. On the other hand, $\tau_t/\tau_s$ diverges as $1/q_0^2|\log q_0|$ for small $q_0$, wherein the screening is very weak. Thus, in the weak screening limit the transport scattering rate is much weaker than the single particle scattering rate.
Note that for 2D graphene \cite{hwang_scattering} we have the ratio $\tau_t/\tau_s \rightarrow  2$ as $q_s \rightarrow \infty$ and $\tau_t/\tau_s \rightarrow 1/q_0$ as $q_0 \rightarrow 0$.   
Thus, when we compare the calculated scattering time ratio of 3D Dirac semimetals with the scattering time ratio of 2D graphene \cite{hwang_scattering} we find that the ratio of 2D graphene is larger than that of the 3D Dirac semimetals in the strong screening limit, but in the weak screening limit the ratio is strongly affected by the interaction parameter in 3D Dirac semimetals. 
This is an unexpected dimensional dependence of the ratio $\tau_t/\tau_s$ in Dirac systems which is not found in the corresponding parabolic band systems.

\begin{figure}
	\centering
	\includegraphics[width=1.\columnwidth]{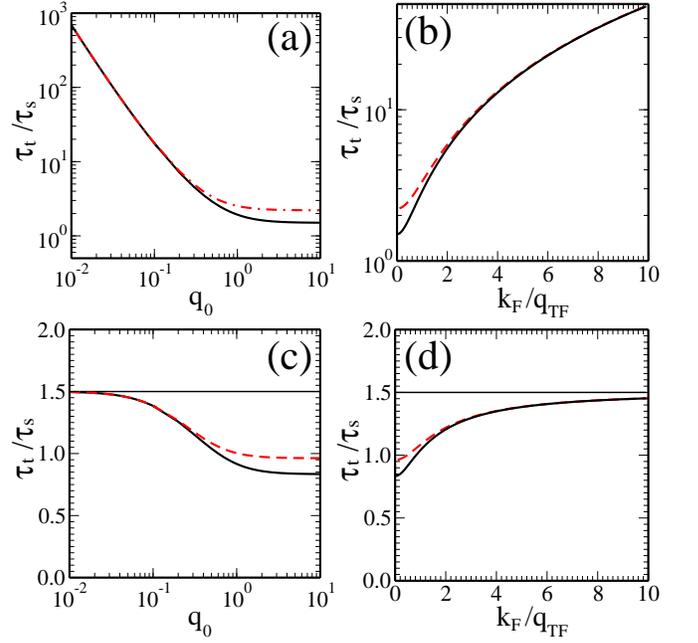}
	\caption{
(a) Calculated ratio of the transport scattering time to the single particle scattering time $\tat/\tas$ as a function of $q_0$ for the screened charged impurity scattering. The solid (dashed) line represents the ratio for charged impurity
with TF (RPA) screening effect. (b) The ratio as a function of $k_F/q_{TF}$. 
Note that $q_0= \sqrt{2g/\pi} \sqrt{\alpha}$, where $\alpha = e^2/\kappa \hbar v_F$ is the dimensionless effective coupling constant. 
(c) and (d) show the scattering time ratio for the short range impurity scattering as a function of $q_0$ and $k_F/q_{TF}$, respectively. The horizontal solid lines in (c) and (d)  indicate the ratio for the $\delta$-range  unscreened short-range disorder. For comparison we also show the results for screened short-range disorders within TF (thick solid lines) and RPA (dashed lines).
}
\label{fig5}
\end{figure}

For the short range disorder with 
$\langle V(\vk,\vk') \rangle = V_0$ we have
\begin{equation}
\frac{1}{\tau_s} = \frac{n_d V_0^2 k_F^2}{2\pi \hbar^2 v_F}.
\label{tauinss}
\end{equation}
Thus, from Eqs.~(\ref{tauins}) and (\ref{tauinss}) we have the ratio $\tau_t/\tau_s = 3/2$, which is the same result we obtain for the very strong screening limit $q_0 \rightarrow \infty$.
Due to the suppression of backward scattering (induced by the  chiral symmetry of 3D Dirac semimetals), the zero-range white-noise disorder potential model  leads to $\tau_t/\tau_s = 3/2$, in contrast to the $\tau_t = \tau_s$ in ordinary metals and semiconductors. This is due to the importance of $k_F$ rather than $2k_F$ scattering in dominating transport properties in 3D Dirac semimetals, whereas the small-angle scattering always dominates $\tau_s$.
A generic observation of $\tau_t > \tau_s$ thus would indicate the Dirac-like behavior of a 3D material in contrast to 3D parabolic systems which tend to have $\tau_t=\tau_s$ in most situations.

Figures~\ref{fig5}(a) and (b) show the ratio of the transport scattering time to the single particle scattering time $\tat/\tas$ of charged impurity scattering as a function of $q_0$ and $k_F/q_{TF}$, respectively. The solid (dashed) line represents the ratio within
TF (RPA) screening. Note that $q_0= \sqrt{2g/\pi} \sqrt{\alpha}$.
For strong screening $q_0 \rightarrow \infty$ (or $k_F/q_{TF} \rightarrow 0$) the ratio approaches the value $3/2$ for TF screening. For RPA screening, $\tau_t/\tau_s > 3/2$ even in the very strong screening limit and is dependent on the ultraviolet cutoff $k_c$. We find $\tau_t/\tau_s$ increases with $k_c$, which arises from the contribution to the large angle scattering ($q \sim k_F$) in the transport scattering time (see Eq.~(\ref{pit0})). 
In Fig.~\ref{fig5} we use $k_c/k_F = 100$. For smaller $k_c/k_F$ (e.g., $k_c/k_F = 10$) we have $\tau_t/\tau_s \sim 3/2$ in the strong screening limit. 
Figs.~\ref{fig5}(c) and (d) show the scattering time ratio for the short range impurity scattering as a function of $q_0$ and $k_F/q_{TF}$, respectively. The horizontal solid lines in (c) and (d) represent the ratio for the $\delta$-range  unscreened short-range disorder and indicate that the ratio is independent of the screening strength. Note that in 2D graphene the ratio for the unscreened $\delta$-range disorder scattering varies with the screening strength \cite{hwang_scattering}. For comparison we also show the results for screened short-range disorders within TF (thick solid lines) and RPA (dashed lines). If the short-range disorder (e.g., atomic point defects) is screened, the single particle relaxation time becomes larger than the transport scattering time in the strong screening limit.  This phenomenon arises from the importance of $k_F$ rather than $2k_F$ scattering in transport properties in 3D Dirac semimetals.

\section{phonon scattering}

In section II we considered the temperature dependent conductivity arising from the temperature dependent screening and energy averaging of the transport scattering time. Since phonons are dominant scattering sources at high temperatures \cite{dassarma2011,andormp}, in this section we consider the longitudinal acoustic (LA) phonons
to calculate the acoustic phonon scattering limited conductivity of the 3D Dirac system. 

When we consider the phonon emission and absorption at finite temperature 
the energy dependent relaxation time [$\tau(\ve_{\vk})$] is defined by \cite{gant}
\begin{equation}
\frac{1}{\tau(\ve_{\vk})} = \sum_{\vk'}(1-\cos\theta_{\vk \vk'}) W_{\vk
  \vk'}\frac{1 - f(\varepsilon')}{1-f(\varepsilon)}
\end{equation}
where $\theta_{\vk \vk'}$ is the scattering angle between $\vk$ and
$\vk'$, $\ve = \hbar v_F |{\vk}|$, and
$W_{\vk \vk'}$ is the transition probability from the state
$\vk$ to $\vk'$, and  $f(\ve)$ is the Fermi distribution function, 
$f(\epsilon_k) =\{ 1+\exp[\beta(\epsilon_k-\mu)] \}^{-1}$ 
with $\beta = 1/k_BT$ and $\mu(T,n)$ is the finite temperature
chemical potential. The chemical potential at finite temperature must be determined  self-consistently to conserve the total number of electrons.
The transition probability can be expressed in terms of the matrix element for scattering \begin{equation}
W_{\vk \vk'}=\frac{2\pi}{\hbar}\sum_{\vq}|C(\vq)|^2
\Delta(\varepsilon,\varepsilon')   
\end{equation}
where $C(\vq)$ is the matrix element for scattering by acoustic phonons
and $\Delta(\ve,\ve')$ is given by
\begin{equation}
\Delta(\ve,\ve') = N_q \delta(\ve-\ve'+\omega_{\vq}) + (N_q + 1)
\delta(\ve-\ve'-\omega_{\vq}),
\label{delta}
\end{equation}
where $\omega_{\vq}=v_{ph} \vq$ is the acoustic phonon energy with
$v_{ph}$ being the phonon velocity  and
$N_q$ is the phonon 
occupation number $N_q = [{\exp(\beta \omega_{\vq}) -1}]^{-1}$.
The first (second) term in Eq.~(\ref{delta}) corresponds to the
absorption (emission) of an acoustic phonon of energy $\omega_{\bf q}$.
The matrix element $C({\bf q})$ is independent of the phonon
occupation numbers and is given by \cite{gant}
\begin{equation}
C({\bf q=k-k'})= \int d^3r \psi_{\bf k'}^*({\bf r}) U({\bf r}) \psi_{\bf k}({\bf r}),
\end{equation}
where $U({\bf r})$ is the potential due to the propagation of phonons and $\psi_{\bf k}({\bf r})$ is the electronic wave function of the 3D Dirac semimetals.
When the chiral symmetry of the 3D Dirac semimetals is considered the matrix element $|C(\vq)|^2$ for the deformation potential is given by
\begin{equation}
|C(\vq)|^2 = \frac{D^2\hbar q}{2V\rho_m v_{ph}} \frac{1+\cos\theta_{\bf kk'}}{2},
\end{equation}
where $D$ is the deformation potential coupling constant
defining the basic electron-phonon interaction strength, $\rho_m$
is the mass density of 3D Dirac semimetals, and $V$ is the volume of the sample.

In this paper we consider the phonon limited resistivity in two distinct transport regimes depending on
whether the phonon system is degenerate (i.e., Bloch-Gr\"{u}neisen, BG) or
non-degenerate (equipartition, EP). The
characteristic temperature $T_{BG}$ is defined as $k_B T_{BG} = 2 \hbar k_F
v_{ph}$, which is given, in 3D Dirac semimetals, by $T_{BG} = 2 v_{ph}k_F/k_B \approx 37.5 \;
v_{ph} n^{1/3}$ K with the phonon velocity measured in unit of $10^6$ cm/s and the density $n$  
measured in unit of $n=10^{18}cm^{-3}$.  
In general, since the Fermi velocity $v_F$ in 3D Dirac semimetals is much larger than the acoustic phonon velocity, the scattering of electrons by acoustic phonons may be considered
quasi-elastically. In the high temperature equipartition regime ($\hbar \omega_{\vq} \ll k_B T$) the relaxation time is calculated to be
\begin{equation}
\frac{1}{\tau(\ve_{\vk})} =
\frac{1}{2\pi \hbar^4}\frac{\ve_{\vk}^2}{3v_F^3}\frac{D^2}{\rho_m v_{ph}^2}
k_BT.
\end{equation}
Thus, in the non-degenerate EP regime the
scattering rate [$1/\tau(\ve_{\vk})$] depends linearly on the
temperature. In particular, when the temperature is lower than the Fermi temperature, i.e.,
$T_{BG} \ll T \ll T_F=E_F/k_B$ we can approximate  $\langle \tau \rangle
\approx \tau(E_F)$. Then, the density dependent scattering time is given as $\langle \tau \rangle
\approx \tau(E_F) \propto n^{-2/3}$. Since the density dependent conductivity is given by $\sigma \propto D_F \langle \tau \rangle \propto n^{2/3} \langle \tau \rangle$,
the calculated phonon limited conductivity is independent of 
the electron density at temperatures much lower than the Fermi temperature.
Note that for 2D graphene we also found the graphene resistivity limited by phonon scattering in EP regime is to be density independent.\cite{hwang_ph} Therefore the any density dependence in the conductivity in the EP regime may entirely arise from the impurity scattering discussed in the Sec. II.
Phonons should not contribute to the density dependence of the 3D Dirac conductivity, except perhaps at very high temperatures.

Finally, we calculate the transport scattering times in the BG regime where
$\hbar \omega_{\vq} \sim k_BT$. 
In the BG regime due to the exponential decrease of phonon population and 
the sharp Fermi distribution, the scattering rate is strongly suppressed.
Thus, the
temperature dependence of the relaxation time via the statistical
occupation factors in Eq. (\ref{delta}) becomes more complicated.
By considering the full occupation factors we have in low temperature limits $T \ll T_{BG}$
we obtain analytically
\begin{equation}
{\langle \tau \rangle}^{-1} \approx  \frac{8}{\pi} \frac{\hbar v_F}{
  E_F^2} \frac{D^2}{2\rho_mv_{ph}}\frac{5! \zeta(5)}{(\hbar
  v_{ph})^5}(k_BT)^5.
\label{tau_t}
\end{equation}
Thus, we find that the temperature dependent resistivity in the BG regime
becomes $\rho \sim T^{5}$ without screening effects. 
We can include the screening effects on the bare scattering rates 
by dividing the matrix elements $C(\vq)$ by the dielectric function of
3D Dirac semimetals, $\epsilon(q) = 1- V(q) \Pi(q)$.
If we include screening effects by the carriers themselves
the low-temperature resistivity goes as $\rho \sim T^{9}$ because the screening function provides an extra $\omega_q^4$ term. For intrinsic 3D Dirac semimetals we have $\rho \sim T^5/[\ln(T)]^2$.
Even though the resistivity in the EP regime is density independent,
Eq. (\ref{tau_t}) indicates that the calculated resistivity in the BG regime
is given by $\rho_{BG} \sim n^{-4/3}$ because $\rho \propto (D_F \langle \tau \rangle)^{-1}$, where $D_F$ is the density of states of 3D Dirac semimetals.
Thus, the BG conductivity limited by phonon scattering at very low temperatures may manifest some density dependence, but it will be difficult to observe it experimentally since the overall phonon scattering effect is suppressed by the factor $T^5$ (or even by $T^9$ if screening is important).  We therefore believe that the low-temperature conductivity of a 3D Dirac system is likely to be dominated by impurity scattering (similar to ordinary semiconductors with phonon effects becoming unobservably small at very low temperatures).  The high temperature conductivity should, however, decrease as $1/T$ due to acoustic phonon scattering as we show here.

\section{discussion and conclusion}

For comparison with experimental results (when they become available),
it is important to emphasize that, unlike in graphene where the
Fermi level and the carrier density can be tuned by externally applied
gate voltages, the only way to obtain a finite carrier density in a 3D
Dirac system is through doping by impurities in which case $n \leq n_i$
will apply generically, leading to the $T=0$ conductivity $\sigma \sim
n^{4/3}/n_i \sim n^{1/3}$ in general, assuming $n=n_i$. Such an
$n^{1/3}$ (or equivalently $n_i^{1/3}$ )
scaling dependence \cite{burkov2011} of the
low-temperature ($T \ll T_F$) conductivity with doping density should
be the hallmark of a 3D doped Dirac system. Since such variable doping
samples are not easy to use in controlled experimental studies (e.g.,
the unintentional background impurities could vary from sample to
sample in an unknown manner), we believe that the study of temperature
scaling of conductivity may be the ideal way of establishing the Dirac
nature of a candidate 3D Dirac system. In particular, the intrinsic
undoped semimetallic behavior should manifest in the high-temperature
limit ($T/T_F \gg 1$), where $\sigma(T) \sim T^4$ if Coulomb disorder
prevails (as is likely to be the case in the presence of charged
dopants or impurities). At low temperatures ($T\ll T_F$), the
extrinsic doped behavior of the conductivity would lead to basically a
temperature independent conductivity.
When we consider  phonon scattering  the resistivity should 
be linear in temperature and density independent in the high temperature equipartition regime ($T_{BG} \ll T$). In the low-temperature BG regime ($T \ll T_{BG}$) the $\rho_{BG}$ shows much higher power law behavior in temperature ($\rho_{BG} \sim T^5$) and the decreasing behavior with density as $\rho_{BG} \sim n^{-4/3}$,
but this is unlikely to be experimentally observable in the near future since very clean systems would be necessary.  Our results for the comparison between transport and single-particle scattering times indicate that the interesting result $\tau_t/\tau_s =3/2$ could apply in a large parameter regime in 3D Dirac systems, which should be experimentally observable.

Actually, we believe that the study of the density-dependent conductivity should be possible in 3D Dirac systems if the carrier density ($n$) itself, rather than the dopant density ($n_i$), is directly measured through the low-field Hall effect.  Then, the conductivity at a fixed low temperature could be plotted against carrier density to obtain the expected $n^{4/3}$ (for Coulomb disorder) or density-independent (point defects or neutral impurities) behavior as a function of carrier density, thus conclusively establishing the Dirac nature of the 3D system.  Since the corresponding non-Dirac parabolic semiconductors have completely different density scaling (see Table 1), the observation of the expected density-scaling in the electrical conductivity should be a smoking gun for the Dirac nature of the 3D materials just as the linear-in-density behavior of conductivity is the hallmark of 2D Dirac systems \cite{dassarma2011}.  We note that the generic density dependence of 3D Dirac conductivity in the presence of both Coulomb and zero-range disorder should be an increasing conductivity as $n^{4/3}$ which should then eventually saturates to a density-independent value at a high (and sample-dependent) carrier density where the short-range disorder effects eventually dominate the conductivity as the Coulomb disorder gets screened out.  Such a crossover from a $n^{4/3}$ density dependence to a density-independent conductivity would be the smoking gun evidence for the Dirac nature of the underlying 3D carrier, particularly if, in addition, one can obtain also the expected $\tau_t > \tau_s$ behavior in the relaxation times as derived in our theory.  We note that such density dependent conductivity studies are carried out routinely in the experimental investigations of metal-insulator transition in 3D doped semiconductors (e.g. P-doped Si \cite{rosenbaum})
where the conductivity scaling as a function of doping density is the key physics being investigated.  In the case of 3D Dirac systems, the great advantage is that one does not need to worry about the localization-induced metal-insulator transition because of the strong chirality-induced suppression of $2k_F$-backscattering, and thus one expects a very large range of doping density (and temperature) over which the expected Drude-Boltzmann density and temperature scaling derived in our work should hold in the actual systems just as it does in the corresponding 2D graphene and 3D topological insulators.

In summary, we have provided a theory for impurity- and phonon-scattering limited
transport properties of 3D Dirac systems, both in undoped and doped
situations and both for high and low temperatures. Our predicted
temperature dependent scaling behavior of the conductivity should
distinguish a Dirac system from an ordinary semiconductor.

\section*{acknoledgement}
This work is supported by LPS-CMTC and NRF-2009-0083540.


\begin{thebibliography}{999}


\bibitem{geim2007} A. K. Geim and K. S. Novoselov, Nat. Mater. {\bf
  6}, 183 (2007).  

\bibitem{neto} C. W. J. Beenakker, Rev. Mod. Phys. {\bf 80}, 1337
  (2008).

\bibitem{peres} N. M. R. Peres, Rev. Mod. Phys. {\bf 82}, 2673 (2010).


\bibitem{dassarma2011}S. Das Sarma, S. Adam, E. H. Hwang, and E. Rossi,
Rev. Mod. Phys. {\bf 83}, 407 (2011). 

\bibitem{tarruell2012} L. Tarruell, D. Greif, T. Uehlinger, G. Jotzu,
  and T. Esslinger, Nature {\bf 483}, 302 (2012).


\bibitem{wan2011} S. M. Young, S. Zaheer, J. C. Y. Teo, C. L. Kane, E. J. Mele,
and A. M. Rappe, Phys. Rev. Lett. {\bf 108}, 140405 (2012); Z. Wang, Y. Sun, X.-Q. Chen, C. Franchini, G. Xu, H. Weng, X. Dai, and Z. Fang, Phys. Rev. B {\bf 85}, 195320 (2012).

\bibitem{singh2012}B. Singh, A. Sharma, H. Lin, M. Z. Hasan, R. Prasad, and
A. Bansil, Phys. Rev. B 86, 115208 (2012): J. C. Smith, S. Banerjee, V. Pardo, and W. E. Pickett, Phys. Rev. Lett. {\bf 106}, 056401 (2011); C.-X. Liu, P. Ye, and X.-L. Qi, Phys. Rev. B {\bf 87}, 235306 (2013);
W. Witczak-Krempa and Y. B. Kim, Phys. Rev. B {\bf 85}, 045124 (2012); G. Xu, H. Weng, Z. Wang, X. Dai, and Z. Fang, Phys. Rev. Lett. {\bf 107}, 186806 (2011).

\bibitem{ohtsuki2014}  K. Kobayashi, T. Ohtsuki, K.-I. Imura, and I. F. Herbut,
Phys. Rev. Lett. {\bf 112}, 016402 (2014); R. Nandkishore, D. A. Huse, and S. Sondhi, arXiv:1307.3252. 

\bibitem{liu2014} Z. K. Liu, B. Zhou, Y. Zhang, Z. J. Wang,
  H. M. Weng, D. Prabhakaran, S.-K. Mo, Z. X. Shen, Z. Fang, X. Dai,
  Z. Hussain, and Y. L. Chen,
Science {\bf 343}, 864 (2014).

\bibitem{borisenko2014} S. Borisenko, Q. Gibson, D. Evtushinsky,
  V. Zabolotnyy, B. B\"{u}chner, and R. J. Cava, Phys. Rev. Lett. {\bf
    113}, 027603 (2014).
\bibitem{neupane} M. Neupane, SuYang Xu, R. Sankar,
N. Alidoust, G. Bian, C. Liu, I. Belopolski, T.-R. Chang,
H.-T. Jeng, H. Lin, A. Bansil, F. Chou, M. Zahid Hasan, arXiv:1406.2318. 


\bibitem{jang2008}C. Jang, S. Adam, J.-H. Chen, E. D. Williams, S. Das
  Sarma, and M. S. Fuhrer, Phys. Rev. Lett. {\bf 101}, 146805 (2008); 
L. A. Ponomarenko, R. Yang, T. M. Mohiuddin, M. I. Katsnelson,
K. S. Novoselov, S. V. Morozov, A. A. Zhukov, F. Schedin, E. W. Hill,
and A. K. Geim, Phys. Rev. Lett. {\bf 102}, 206603 (2009);
A. K. M. Newaz, Y. S. Puzyrev, B. Wang, S. T. Pantelides, and
K. I. Bolotin, Nat. Commun. {\bf 3}, 1740 (2012). 


\bibitem{ti} Dimitrie Culcer, E. H. Hwang, Tudor D. Stanescu, and S. Das Sarma, Phys. Rev. B {\bf 82}, 155457 (2010); S. Adam, E. H. Hwang, S. Das Sarma,
Phys. Rev. B {\bf 85}, 235413 (2012); D. Kim, P. Syers, N. P. Butch, J. Paglione, and M. S. Fuhrer Nat Comms {\bf 4}, 2040 (2013);
D. Kong, Y. Chen, J. J. Cha, Q. Zhang, J. G. Analytis, K. Lai, Z. Liu, S. S. Hong, K. J. Koski, S. K. Mo, Z. Hussain, I. R. Fisher, Z. X. Shen, and Y. Cui, Nature Nano. {\bf 6}, 705 (2011).


\bibitem{hosur2012} P. Hosur, S. A. Parameswaran, and A. Vishwanath, Phys. Rev. Lett. {\bf 108}, 046602 (2012).

\bibitem{sbierski2012} B. Sbierski, G. Pohl, E. J. Bergholtz, and P. W. Brouwer, Phys. Rev. Lett. {\bf 113}, 026602 (2014).

\bibitem{skinner}  B. Skinner, arXiv:1406.2318 (2014).

\bibitem{wan2001} X. Wan, A. M. Turner, A. Vishwanath, and S. Y. Savrasov, Phys.
Rev. B {\bf 83}, 205101 (2011).

\bibitem{biswas2014} R. R. Biswas and S. Ryu, Phys. Rev. B {\bf 89}, 014205 (2014).


\bibitem{hwang2007} E. H. Hwang, S. Adam, and S. Das Sarma,
Phys. Rev. Lett. {\bf 98}, 186806 (2007).


\bibitem{tan2007} Y.-W. Tan, Y. Zhang, K. Bolotin, Y. Zhao, S. Adam,
  E. H. Hwang, S. Das Sarma, H. L. St\"{o}rmer, and P. Kim, 
Phys. Rev. Lett. {\bf 99}, 246803 (2007). 

\bibitem{adam2007} S. Adam, E. H. Hwang, V. Galitski, and S. Das
  Sarma, Proc. Natl Acad. Sci. {\bf 104}, 18392 (2007).

\bibitem{chen2008} J.-H. Chen, C. Jang, S. Adam,
  M. S. Fuhrer, E. D. Williams, and M. Ishigami, Nat. Phys. {\bf 4},
  377 (2008). 


\bibitem{andormp} T. Ando, A. B. Fowler, and F. Stern,
  Rev. Mod. Phys. {\bf 54},  437 (1982). 


\bibitem{mahan} G. D. Mahan, {\it Many Particle Physics} (Plenum, New York, 1993).

\bibitem{gant} V. F. Gantmakher and Y. B. Levinson, {\it Carrier Scattering in Metals and Semiconductors} (Elsevier, Amsterdam, 1987).

\bibitem{dassarma1985} S. Das Sarma and F. Stern, Phys. Rev. B {\bf 32}, 8442 (1985).

\bibitem{hwang_scattering} E. H. Hwang and S. Das Sarma, Phys. Rev. B {\bf 77}, 195412 (2008).

\bibitem{2Dscreening}S. Das Sarma, Phys. Rev. B {\bf 33}, 5401 (1986).

\bibitem{lindhart} J. Lindhard, K. Dan. Vidensk. Selsk., Mat.-Fys. Medd. {\bf 28}, 8 (1954).


\bibitem{screening} E. H. Hwang and S. Das Sarma,  Phys. Rev. B {\bf 75}, 205418 (2007);
Phys. Rev. B {\bf 79}, 165404 (2009).


\bibitem{dassarma_scaling} S. Das Sarma and E. H. Hwang, Phys. Rev. B {\bf 88}, 035439 (2013).

\bibitem{hwang_ph} E. H. Hwang and S. Das Sarma Phys. Rev. B {\bf 77}, 115449 (2008).

\bibitem{burkov2011} A. A. Burkov, M. D. Hook, and Leon Balents,
  Phys. Rev. B {\bf 84}, 235126 (2011).

\bibitem{rosenbaum}T. F. Rosenbaum, R. F. Milligan, M. A. Paalanen, G. A. Thomas, R. N. Bhatt, and W. Lin, Phys. Rev. B {\bf 27}, 7509 (1983).

\end{thebibliography}
\end{document}